\documentclass[%
 reprint,
nofootinbib,
nobibnotes,
bibnotes,graphicx,graphics,
 amsmath,amssymb,
 aps,stmaryrd,
]{revtex4-1}
\usepackage{natbib} 
\usepackage{soul}
\newcommand{\comment}[1]{}
\usepackage{tikz}
  \newlength\squareheight
\usepackage{wasysym} 
\usepackage{graphicx}
\usepackage{dcolumn}
\usepackage{bm}

\begin{document}

\preprint{Draft}

\title{Cooperative coinfection dynamics on clustered networks}

\author{Peter Mann}
\email{pm78@st-andrews.ac.uk}
\author{V. Anne Smith}%
\author{John B.O. Mitchell}
\author{Simon Dobson}
\affiliation{School of Computer Science, University of St Andrews, St Andrews, Fife KY16 9SX, United Kingdom }
\affiliation{School of Chemistry, University of St Andrews, St Andrews, Fife KY16 9ST, United Kingdom }
\affiliation{School of Biology, University of St Andrews, St Andrews, Fife KY16 9TH, United Kingdom }

\date{\today}

\begin{abstract}
Coinfection is the process by which a host that is infected with a pathogen becomes infected by a second pathogen at a later point in time. An immunosuppressant host response to a primary disease can facilitate spreading of a subsequent emergent pathogen among the population. Social contact patterns within the substrate populace can be modelled using complex networks and it has been shown that contact patterns vastly influence the emergent disease dynamics. In this paper, we consider the effect of contact clustering on the coinfection dynamics of two pathogens spreading over a network. We use the generating function formulation to describe the expected outbreak sizes of each pathogen and numerically study the threshold criteria that permit the coexistence of each strain among the network. We find that the effects of clustering on the levels of coinfection are governed by the details of the contact topology.
\end{abstract}

\pacs{Valid PACS appear here}
\maketitle



Coinfection is the process by which a host that is infected by a single pathogen becomes further infected by a second pathogen after a period of time. Clinically this is of great concern, as coinfection can worsen human health when compared to mono-infections. Well known examples of coinfection occur in HIV infected hosts that subsequently become infected with Hepatitis B or C, hosts with syphilis or HSV-2 that later contract HIV or the increase in tuberculosis cases during the 1918-1919 Spanish flu. \cite{PMID:15219556,Mavilia2018,DEMONTE201627,Brundage2008}. 

At the time of writing this article the SARS-CoV-2 pandemic is continuing to spread around the globe. The genetics of SARS-CoV-2 is not fixed, however. The virus can evolve through a number of means, including transcription errors and the exchange of plasmids with infections co-existing in a host. Both of these mechanisms potentially - in a very few cases - lead to mutations that are advantageous to the virus' lifecycle or spread. The larger the epidemic is, and the longer it continues, the more opportunities there are for such mutations to arise. We must therefore consider, with urgency, the impact that the emergence of a second strain would have on a population. This is not simply a question of biology: since the population is currently the subject of various epidemic management strategies, we must also consider the role of contact structure on the emergence and propagation of any novel strain. Chief among these strategies is \textit{social distancing}, whereby tight clusters of contacts are discouraged and instead a sparser, less dense system of contacts is formed, slowing down disease transmission.   

Contact structures can be modelled by complex networks: the nodes representing hosts and edges representing their contacts which are sufficient for disease transmission. The governing topology of the contact network has a profound influence on the threshold at which a localised disease outbreak becomes an epidemic over the entire network, as well as on its final size. Contact clustering occurs when the edges of any three nodes are connected in a triangle arrangement. It is known that the effects of clustering on the outbreak size of a disease are highly dependent on the manner in which the triangles are present in the network. Newman \cite{PhysRevLett.103.058701} found that for Poisson graphs, clustering reduced the epidemic threshold and decreased the overall outbreak size of a disease. Conversely, Miller \cite{miller_2009,10.1371/journal.pcbi.1002042,citekey2} showed that the decrease in the epidemic threshold was a nuance of the degree correlations within the Poisson random graph model. Within the outbreak of the first strain, these correlations play the dominant role in the emergent properties. When considering the degree-$\delta$ model proposed by Hackett \textit{et alia} \cite{PhysRevE.83.056107}, which affords control of the assortativity among the degree correlations, Miller found that clustering had a dual effect depending on precisely how the networks are constructed. In contrast to the Poisson model, clustering can \textit{increase} the epidemic threshold for $\delta$-model networks. This was also supported by Gleeson \textit{et alia} \cite{PhysRevE.81.066114} for $\gamma$-theory networks. These dichotomous findings are insight into the difficulty in controlling a single aspect of network behaviour when modulating substrate topology. Recent studies by Hasegawa and Mizutaka \cite{PhysRevE.101.062310,Mizutaka_2020} indicate that Poisson networks display dissasortative degree correlations at the critical point, whilst degree-$\delta$ models can be made to display positive assortativity, especially among low degree nodes. The positive correlations were shown to vanish upon renormalisation to the characteristic scale of a triangle, however.

Bond percolation is a stochastic process that iterates the edges of a network and \textit{occupies} them with probability $T$, or fails to occupy them with probability $1-T$. The emergence of a giant connected component (GCC) among the occupied edges in the network occurs through a second order phase transition as a function of increasing bond occupancy probability. Bond occupation over networks is well known to represent epidemic outbreaks among populations \cite{PhysRevE.66.016128,PhysRevE.76.036113}; bond percolation probability representing the transmissibility of the disease and the GCC indicating the expected outbreak size. 

In a recent paper \cite{2020arXiv200703287M}, we studied the effect of contact clustering on the spread of two sequential pathogens that interact via a complete cross-immune coupling using bond percolation. In our previous model, infection with strain-1 prevented any further infections and only uninfected or singly-infected hosts were found in the absorbing state of the model. In this paper, we study the antithesis of that model, namely, the scenario where infection with strain-1 is an absolute \textit{prerequisite} for infection with a second strain; a perfect coinfection model. Such an essential condition on the infection of a host with a second strain is the limiting behavior of a more general disease interaction mechanism, \textit{partial immunity}, whereby a previous strain only modulates the probability of infection by subsequent pathogens \cite{BANSAL2012176}. 

Dynamics in which the spread of a disease is facilitated and actively promoted by an earlier disease is said to exhibit cooperative spreading behavior. Cooperation has been studied before in the context of concurrent diseases and networks \cite{Chen_2013,10.1371/journal.pone.0071321}; however, these studies focused on locally tree-like network structures, where the density of cycles falls to zero as the size of the network becomes infinite. Real-world contact topologies often contain closed loops, since, any two neighbors of a given node are often mutually connected together. Recent studies \cite{PhysRevResearch.2.033306,Hebert-Dufresne10551,PhysRevE.99.022301} have been performed on contemporaneous cooperative diseases in the presence of clustered contact topologies; however, to our knowledge, a bond percolation model of coinfection has not yet been proposed for clustered network. In this paper, we use the generating function formulation (GFF) \cite{PhysRevE.64.026118,PhysRevE.66.016128,PhysRevResearch.2.033306} to describe the coinfection dynamics that occur when two pathogens, that are temporally separated,  cooperatively spread over a contact network that has non-trivial clustering. 

\section{Configuration model networks and generating functions}
\label{sec:Background}

The archetypal network model is the Erd\H{o}s-R\'enyi (ER) random graph, a member of the exponential random graph ensemble. This model has received enormous attention in the network science literature in part due to its ability to exhibit the salient features of a network study and in part due to its simplicity. In the ER model, any two nodes within a network of $N$ vertices are connected with probability $\varphi\approx \langle k \rangle/N$, where $\langle k \rangle$ is the average degree of a vertex. In the limit of large network sizes and sparse connectivity, the ER model is well represented by a Poisson distribution. To see this, we use the GFF \cite{PhysRevE.64.026118,PhysRevE.66.016128}. The GFF rests upon the degree distribution, $p(k)$, the probability of choosing a node at random from the network of degree $k$. The degree distribution for ER networks is 
\begin{equation}
    p(k)=\binom{N}{k}\varphi^k(1-\varphi)^{N-k}\simeq \frac{\langle k\rangle^ke^{-\langle k\rangle}}{k!}
\end{equation}
where $\langle k \rangle$ is the average degree. We recognise this as a Poisson distribution. 

When the network contains triangles, we introduce the joint degree distribution, $p(s , t)$, which is the probability of choosing a node at random from the network with $s $ tree-like edges and $t$ triangles \cite{miller_2009,PhysRevLett.103.058701}. We can recover $p(k)$ from the joint degree sequence as 
\begin{equation}
    p(k) = \sum^\infty_{s =0}\sum^\infty_{t=0}p(s ,t)\delta_{k,s  + 2t}
\end{equation}
where $\delta_{i,j}$ is the Kronecker delta. The joint probability distribution is generated by a bivariate generating function as
\begin{equation}
    G_0(z_\bot,z_\Delta) = \sum^\infty_{s =0}\sum^\infty_{t=0} p(s ,t){z_\bot}^{s }{z_\Delta}^{t}\label{eq:G0_generic}
\end{equation}
The probability of reaching a node of joint degree $(s ,t)$ by following a random tree-like edge back to a node is generated by 
\begin{equation}
    G_{1,\bot}(z_\bot,z_\Delta) = \frac{1}{\langle s  \rangle}\frac{\partial G_0}{\partial z_\bot}
\end{equation}
Similarly, the degree of the node reached by following a random triangle edge to a node is 
\begin{equation}
    G_{1,\Delta}(z_\bot,z_\Delta) = \frac{1}{\langle t \rangle}\frac{\partial G_0}{\partial z_\Delta}
\end{equation}
In each case, $\langle s\rangle$ is the average number of tree-like edges a node is a member of, and is given by $\partial_{z_s} G_0(1,1)$, and similarly for $\langle t\rangle$. 

The clustering coefficient $C$ is a metric that indicates the level of clustering in the network \cite{PhysRevLett.103.058701, PhysRevE.74.056114} by measuring how many connected triples are closed into a cycle. It is given by the following quotient
\begin{equation}
    C = \frac{3N_\Delta}{N_3}\label{eq:clusteringCoeff}
\end{equation}
where $N_\Delta$ is the number of triangles and $N_3$ is the number of connected triples \cite{PhysRevLett.103.058701}. With these definitions we can proceed to study the percolation properties of clustered networks. 
    
    The random graphs we consider here are built using the generalised configuration model \cite{PhysRevE.82.066118,2020arXiv200606744M,mann2020random}. In this model, each node is assigned a stub-degree, $s$ and $t$, representing the number of tree-like and the number of triangles a node belongs to, distributed according to $p(s,t)$. For instance, a degree $k=5$ node involved in 3 tree-like edges and 1 triangle has $s = 3$ and $t=1$. Once all nodes have a stub degree, the stubs are connected together by choosing random partners to create a realisation of a random graph whose edge topologies are distributed according to the assigned stub-degree. In this construction, the accidental formation of higher-order cycles is vanishingly small as the network size diverges.
    
\section{Sequential strain model with clustering}
\label{sec:sec1}
In this section we introduce the two-strain model on clustered networks containing triangles and tree-like edges. In this model, we assume that the second strain occurs after the dynamics of the first has equilibrated.

\subsection{Strain-1}\label{subsec:s1}

Once the dynamics of the first pathogen have run their course over the network, the nodes have either been infected or remain uninfected; a binary state equilibrium. To study an infected node in the GCC of strain-1, we must examine all the permissible final states that could surround the node through each edge type and assign a probability to each one. We can then sum the combinations of each state by creating a generating function that encapsulates the total probability of finding a particular infected node with a given final-state neighbour distribution. In this way, we use the local environment of the node to average over all possible neighbour states, weighted by the degree distribution, and then build a macroscopic description of the percolation properties of the entire network.
\begin{figure}[ht!]
\begin{center}
\includegraphics[width=0.33\textwidth]{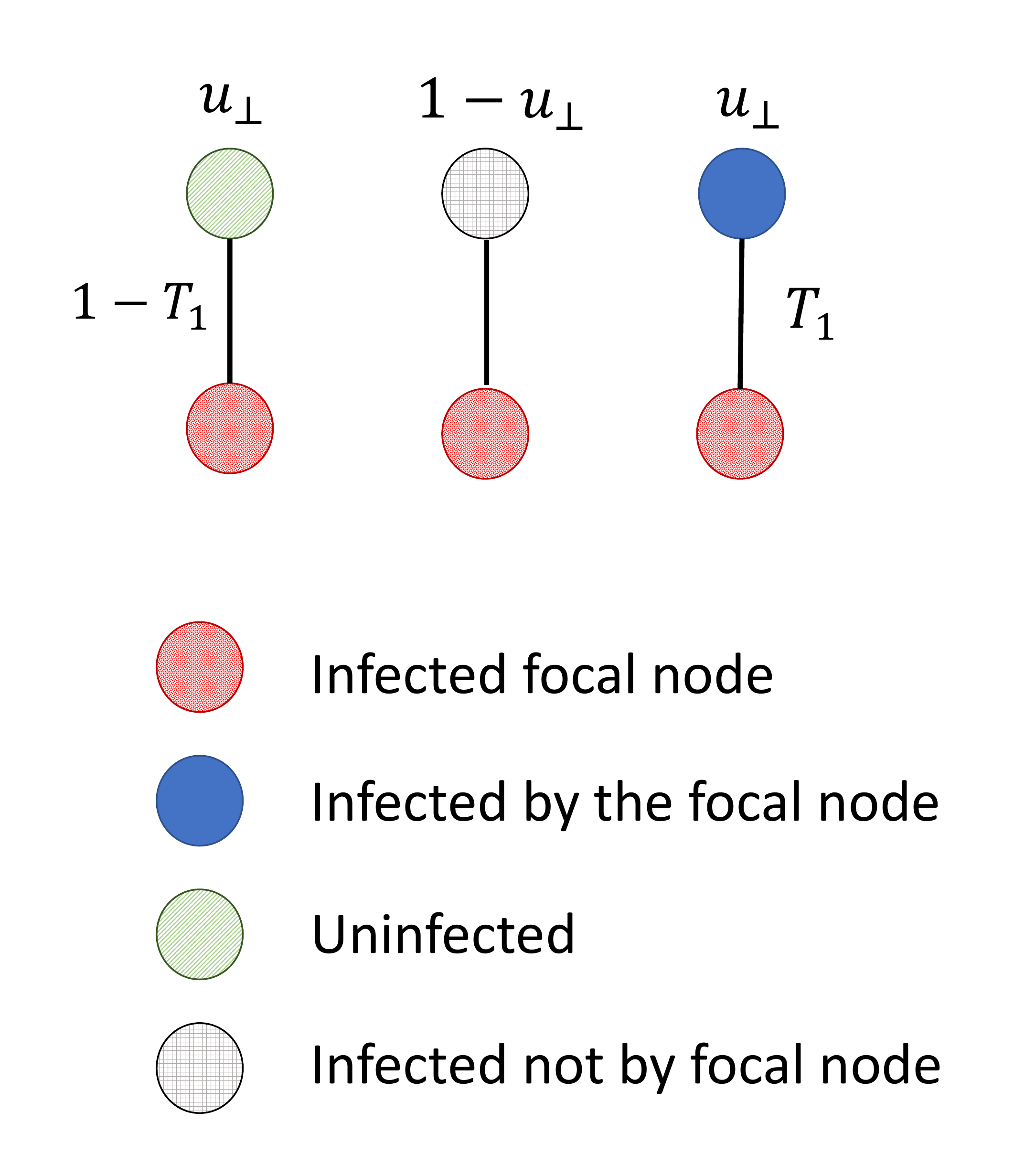}
\caption[treelike]{ 
The tree-like edge topologies found in the GCC following the first strain and their probabilities.
} \label{fig:treelike}
\end{center}
\end{figure}
To do this we must find the probability that the infected node transmitted or failed to transmit its infection to a neighbour during the dynamics of strain-1 through each topological edge-type. We do this first for tree-like edges as they are simpler than triangles and we note that a similar formula was found by Newman \textit{et alia} in \cite{10.1371/journal.pone.0071321}. 

Assuming that the focal node is infected, there are three kinds of tree-like neighbours we can expect after strain-1: uninfected, infected (not by the focal node) and infected (by the focal node directly) according to Fig \ref{fig:treelike}. Defining $u_\bot$ to be the probability that the neighbour found by following a tree-like edge was uninfected, the probability that it doesn't then become infected by the focal node is $1-T_1$. The probability that the neighbour was infected by nodes other than the focal node is simply $1-u_\bot$. Finally, nodes that were uninfected by their other neighbours can be infected directly by the focal node with probability $u_\bot T_1$. 

Therefore, an infected node with $s $ tree-like neighbours, of which $l$ remain uninfected, $m$ are infected by their neighbours (other than the focal node) and $m'=s -l-m$ are infected directly by the focal node, occurs with the following probability
\begin{align}
    f_\bot(u_\bot;T_1) =\ & \binom{s }{l} [u_\bot(1-T_1)]^l]\binom{s -l}{m}\nonumber\\
    &\times[1-u_\bot]^m [u_\bot T_1]^{s -l-m}\nonumber\\
    &\times [1-(1-T_1)^m]\label{eq:f1tree}
\end{align}
The terminal bracket accounts for the probability that one of the $m$ neighbours \textit{must} have infected the focal node. This is expressed as one minus the probability that all $m$ neighbours fail to infect it, each failure occurring with probability $1-T_1$.

\begin{figure}[ht!]
\begin{center}
\includegraphics[width=0.42\textwidth]{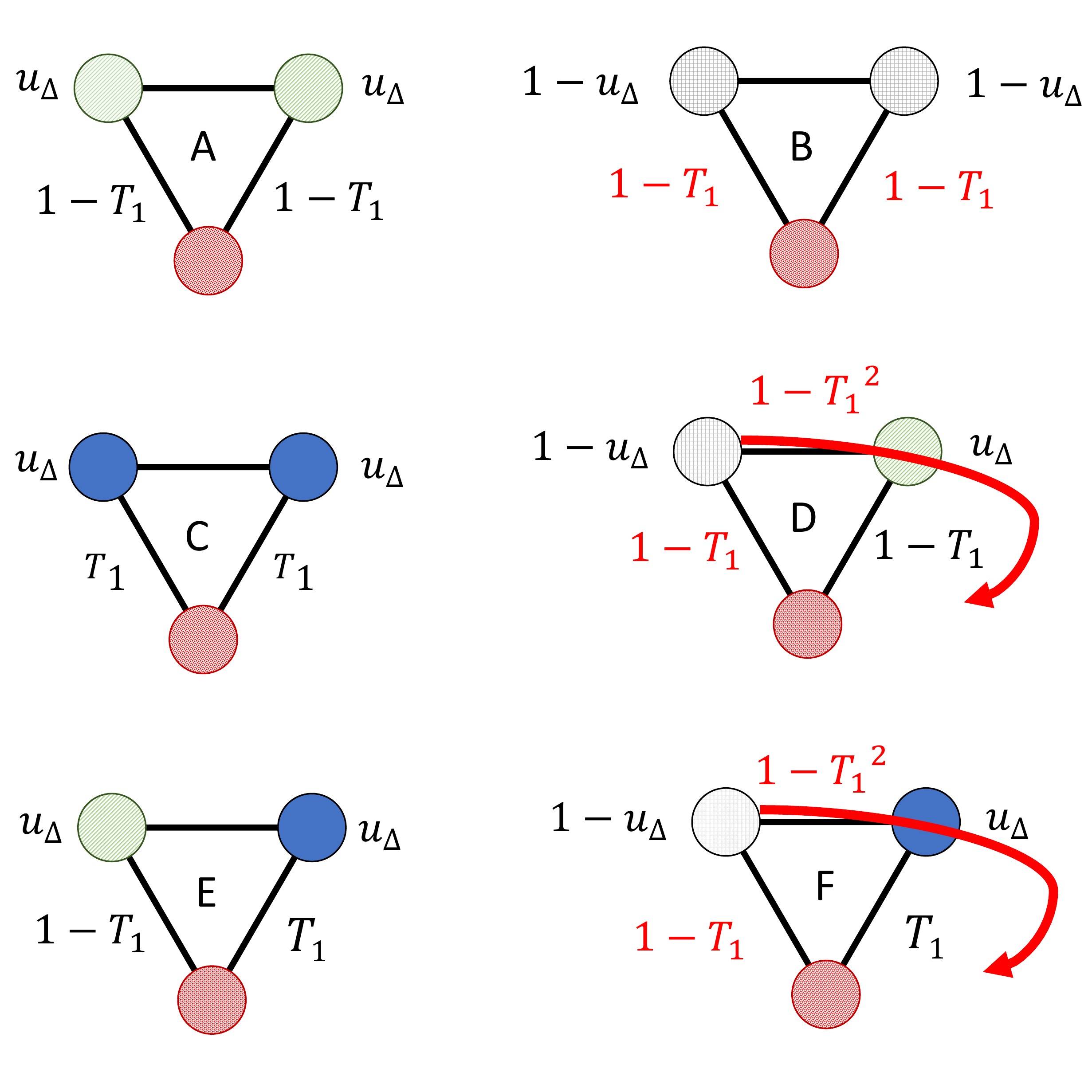}
\caption[triangles]{ 
The 3-cycle edge-topologies found in the GCC following the first strain and their probabilities. Node colours and patters are defined according to Fig \ref{fig:treelike}. Triangles D, E and F consist of inhomogeneous neighbour-states and hence, due to the symmetry of the shape, their reflection about a vertical axis bisecting the focal node can also occur with equal probability. The curved arrows in triangles D and F indicate the additional pathway through the cycle that the infected neighbour could infect the focal node.  
} \label{fig:triangles}
\end{center}
\end{figure}

The corresponding equation for triangles, $f_\Delta(u_\Delta;T_1)$ is a more involved calculation which we now examine. Defining $u_\Delta$ to be the probability that a node involved in a triangle is uninfected, there are six basis triangles to consider following strain-1, see Fig \ref{fig:triangles}. 

We will now discuss each triangle in turn from Fig \ref{fig:triangles}. Triangle A assumes that both neighbours are uninfected, each with probability $u_\Delta(1-T_1)$. Triangle B assumes that each neighbour is infected by means other than the focal node, each occur with probability $1-u_\Delta$. Similarly to the tree-like edge-topology, the focal node infects a neighbour with probability $u_\Delta T_1$, this occurs twice in triangle C. The remaining triangles D,E and F can also be formed by swapping each neighbour with equal probability of occurrence, hence, these configurations contribute twice to the neighbour-state distribution.

With these considerations in mind, the probability that a focal node involved in $t $ triangles having precisely $a$ of type A, $b$ of type B (an so on) is 
\begin{widetext}
\begin{align}
    f_\Delta(u_\Delta;T_1) =\ & \binom{t }{a}[u_\Delta(1-T_1)]^{2a}\binom{t -a}{b}[1-u_\Delta]^{2b}\binom{t -a-b}{c}[u_\Delta T_1]^{2c}\binom{t -a-b-c}{d}[2u_\Delta(1-u_\Delta)(1-T_1)]^d\nonumber\\
    &\times\binom{t -a-b-c-d}{e}[2u_\Delta^2T_1(1-T_1)]^e[2u_\Delta(1-u_\Delta)T_1]^{t -a-b-c-d-e}[1-(1-T_1)^{2b+d+e}(1-T_1^2)^{d+e}]\label{eq:f1triangle}
\end{align}
The terminal bracket accounts for the total probability that one of the infected neighbours (other than those the focal node infected) actually managed to transmit the infection to the focal node in the first place. This probability is constructed as one minus the probability that all the previously infected nodes failed to transmit their infection. Transmission can fail to occur in two ways in the 3-cycle: either directly with probability $1-T_1$, or around the cycle in the special case that the adjoining neighbour was initially uninfected, which occurs with probability $1-T_1^2$. Both methods are highlighted in red in Fig \ref{fig:triangles}. Cycle B has two direct edges and cycles D and E are free to transmit around the outer skeleton of the triangle prior to the infection of the focal node.

Due to the terminal brackets in Eqs \ref{eq:f1tree} and \ref{eq:f1triangle}, the generating functions consist of two terms, the first considers all of the infections that all infected nodes create and amounts to unity, while the second subtracts those that were not part of the GCC or analogously the major outbreak of the strain due to the failure of the indirectly infected neighbours to infect the focal node. 

To construct the generating functions we must insert both $f_\bot$ and $f_\Delta$ into Eq \ref{eq:G0_generic} and sum over each index 
\begin{align}
    H_0(\vec x) =\ &\sum_{s =0}^\infty\sum_{t=0}^\infty p(s ,t) f_\bot f_\Delta x_1^m x_2^{s -l-m}x_3^{2b}x_4^{2c}x_5^dx_6^ex_7^{\lambda-d-e}
    \\
    H_0(\vec x) =\ &\sum_{s =0}^\infty\sum_{t=0}^\infty p(s ,t)\binom{s }{l}[u_\bot(1-T_1)]^l\binom{s -l}{m}[(1-u_\bot)x_1]^m[u_\bot T_1x_2]^{s -l-m}\sum_{a=0}^t  \binom{t }{a}[u_\Delta(1-T_1)]^{2a}\nonumber\\
    &\times\sum_{b=0}^{t -a}\binom{t -a}{b}[(1-u_\Delta)x_3]^{2b}\sum_{d=0}^{t -a-b-c}\binom{t -a-b}{c}[u_\Delta T_1x_4]^{2c}\sum_{d=0}^{\lambda}\binom{\lambda}{d}[2u_\Delta(1-u_\Delta)(1-T_1)x_5]^d\nonumber\\
    &\times\sum_{e=0}^{\lambda-d}\binom{\lambda-d}{e}[2u_\Delta^2T_1(1-T_1)x_6]^e[2u_\Delta(1-u_\Delta)T_1x_7]^{\lambda-d-e}[1-(1-T_1)^{m +2b+d+e}(1-T_1^2)^{d+e}]\label{eq:f1triangleWorking1}
\end{align}
with $\lambda=t -a-b-c$. Applying the binomial theorem we obtain
\begin{align}
    H_0(\vec x)=\ &
    G_0(u_\bot(1-T_1)+(1-u_\bot)x_1 + u_\bot T_1x_2,u_\Delta^2(1-T_1)^2+((1-u_\Delta)x_3)^2+(u_\Delta T_1x_4)^2 \nonumber\\
    &+ 2u_\Delta(1-T_1)(1-u_\Delta)x_5+2u_\Delta(1-T_1)u_\Delta T_1x_6 + 2(1-u_\Delta)u_\Delta T_1x_7)\nonumber\\
    &  - G_0(u_\bot(1-T_1)+(1-u_\bot)(1-T_1)x_1+ u_\bot T_1x_2,u_\Delta^2(1-T_1)^2+((1-u_\Delta)(1-T_1)x_3)^2+(u_\Delta T_1x_4)^2 \nonumber\\
    & +2u_\Delta(1-T_1)^2(1-u_\Delta)(1-T_1^2)x_5+2u_\Delta(1-T_1)u_\Delta T_1x_6 + 2(1-u_\Delta)u_\Delta T_1(1-T_1)(1-T_1^2)x_7)\label{eq:H0}
\end{align}
We will also define $H_{1,\tau}$ as the tautological analogue of Eq \ref{eq:H0}; however, in this case the $G_0$ generating function is replaced by $G_{1,\tau}$. Each $u_\tau$ value then satisfies a self-consistent equation given by 
\begin{equation}
    u_\tau = J_{1,\tau}(\vec x),\qquad \vec x = \{1,\dots, 1\}\label{eq:u1valuesG1}
\end{equation}
where
\begin{align}
    J_{1,\tau}(\vec x) =\ & G_{1,\tau}(u_\bot(1-T_1)+(1-u_\bot)(1-T_1)x_1+ u_\bot T_1x_2,u_\Delta^2(1-T_1)^2+((1-u_\Delta)(1-T_1)x_3)^2+(u_\Delta T_1x_4)^2 \nonumber\\
    &+2u_\Delta(1-T_1)^2(1-u_\Delta)(1-T_1^2)x_5+2u_\Delta(1-T_1)u_\Delta T_1x_6 +2(1-u_\Delta)u_\Delta T_1(1-T_1)(1-T_1^2)x_7)\label{eq:disease1G_1}
\end{align}
which is the argument of the second bracket of Eq \ref{eq:H0}. The outbreak size of the first pathogen, $S_1$, can be generated by the following expression 
\begin{equation}
    S_1[u_\bot,u_\Delta;T_1] = H_0(\vec x), \qquad  \{x = 1, \ \forall x \in \vec x\}\label{eq:disease1G_0}
\end{equation}
\end{widetext}

\subsection{Strain-2}
\label{subsec:s2}

When the second pathogen emerges on the network, the immunological landscape it experiences consists of nodes that became infected by the first disease and nodes that remained uninfected. An additional consideration is the infection history of each infected neighbour; they could have received the disease from the focal node itself or via other nodes they are connected to. To retain this important epidemiological information, it is necessary to define multiple probabilities, $v_\tau$, of not becoming infected by the second disease for each scenario present in the GCC following the first pathogen. In other words, there is a $v_\tau$ value for each distinct node-site in Figs \ref{fig:treelike} and \ref{fig:triangles} that include disease-1 infected nodes. This is in analogy to the analysis in section \ref{subsec:s1} that defined a $u_\tau$ value for each subgraph that the first strain is incident upon in the contact network. In more detail, cycles B, C and F retain their triangle topology in the GCC of strain-1; cycles D and E have become fractured by strain-1 and hence spread strain-2 as if they were in fact trees; finally, the two neighbours in cycle A do not allow the proliferation of strain-2 under the limit of perfect coinfection because they both remain susceptible following the first epidemic. Given the topologies above, we determine that there are eight distinct node-sites and hence eight $v_\tau$ values required to define the second pathogen. These include: two tree-like values $v_\bot^A$ and $v_\bot^B$ for the externally- and directly-infected neighbours in Fig \ref{fig:treelike}, respectively; along with $v_\Delta^B$, $v_\Delta^C$, $v_\Delta^D$, $v_\Delta^E$ and $v_\Delta^{F1}$ and $v_\Delta^{F2}$ for each triangle in Fig \ref{fig:triangles} that has neighbours in the GCC of disease-1. Cycle F contains two infected nodes; however, their infection histories are non-equivalent, each requiring a unique description. 

The probability, $S_2$, that the focal node does not contract disease-2, given that it had contracted disease-1 is the found to be \begin{align}
    S_2= H_0(g(v_\bot^A),g(v_\bot^B),h(v_\Delta^B),h(v_\Delta^C),\nonumber
    \\g(v_\Delta^D),g(v_\Delta^E),h(v_\Delta^{F_1},v_\Delta^{F_2}))/S_1\label{eq:disease2G0}
\end{align}
where $g(v) = v +(1-v)(1-T_2)$ is the probability that a single edge fails to transmit disease-2; and 
\begin{align}
    h(v_1,v_2) =\ & g(v_1)g(v_2)\nonumber\\
    &- (v_1+v_2-2v_1v_2)T_2^2(1-T_2)
\end{align}
with the notation convention that $h(v,v):=h(v)$, is the probability that infection fails when a node belongs to a triangle. The interpretation of Eq \ref{eq:disease2G0} is that the first bracket calculates the spreading of the second disease over the infected subgraph from which we then subtract those contributions that were not part of the GCC, or the major outbreak, of the network. 
It remains now to compute the 8 probabilities $v_\tau^\alpha$ defined above. The recipe for these is quite straightforward: we compute the probability that each site fails to infect the focal node with disease-2, given that the focal node did indeed have disease-1. The $H_{1,\tau}(\vec x)$ generating function can then be used to derive some of the probabilities that a neighbouring node belonging to a given site fails to infect the focal node. Within these 8 scenario probabilities, we must correctly normalise by the probability of obtaining a particular site as well as multiplying by the probability that the focal node did or did not infect it. We find that 
\begin{subequations}
\begin{align}
    v_\bot^A =H_{1,\bot}(g(v_\bot^A),g(v_\bot^B),h(v_\Delta^B),h(v_\Delta^C),\nonumber
    \\g(v_\Delta^D),g(v_\Delta^E),h(v_\Delta^{F_1},v_\Delta^{F_2}))/(1-u_\bot)\label{eq:vbotA}
\end{align}
\begin{align}
    v_\bot^B =J_{1,\bot}(g(v_\bot^A),g(v_\bot^B),h(v_\Delta^B),h(v_\Delta^C),\nonumber
    \\g(v_\Delta^D),g(v_\Delta^E),h(v_\Delta^{F_1},v_\Delta^{F_2}))/u_\bot\label{eq:vbotB}
\end{align}
\begin{align}
    v_\Delta^B =H_{1,\Delta}(g(v_\bot^A),g(v_\bot^B),h(v_\Delta^B),h(v_\Delta^C),\nonumber
    \\g(v_\Delta^D),g(v_\Delta^E),h(v_\Delta^{F_1},v_\Delta^{F_2}))/(1-u_\Delta)\label{eq:vDeltaB}
\end{align}
\begin{align}
    v_\Delta^C =J_{1,\Delta}(g(v_\bot^A),g(v_\bot^B),h(v_\Delta^B),h(v_\Delta^C),\nonumber
    \\g(v_\Delta^D),g(v_\Delta^E),h(v_\Delta^{F_1},v_\Delta^{F_2}))/u_\Delta\label{eq:vDeltaC}
\end{align}
\begin{align}
    v_\Delta^D =H_{1,\Delta}(g(v_\bot^A),g(v_\bot^B),h(v_\Delta^B),h(v_\Delta^C),\nonumber
    \\g(v_\Delta^D),g(v_\Delta^E),h(v_\Delta^{F_1},v_\Delta^{F_2}))/(1-u_\Delta)\label{eq:vDeltaD}
\end{align}
\begin{align}
    v_\Delta^E =J_{1,\Delta}(g(v_\bot^A),g(v_\bot^B),h(v_\Delta^B),h(v_\Delta^C),\nonumber
    \\g(v_\Delta^D),g(v_\Delta^E),h(v_\Delta^{F_1},v_\Delta^{F_2}))/u_\Delta\label{eq:vDeltaE}
\end{align}
\begin{align}
    v_\Delta^{F_1} =H_{1,\Delta}(g(v_\bot^A),g(v_\bot^B),h(v_\Delta^B),h(v_\Delta^C),\nonumber
    \\g(v_\Delta^D),g(v_\Delta^E),h(v_\Delta^{F_1},v_\Delta^{F_2}))/(1-u_\Delta)\label{eq:vDeltaF1}
\end{align}
\begin{align}
    v_\Delta^{F_2} =J_{1,\Delta}(g(v_\bot^A),g(v_\bot^B),h(v_\Delta^B),h(v_\Delta^C),\nonumber
    \\g(v_\Delta^D),g(v_\Delta^E),h(v_\Delta^{F_1},v_\Delta^{F_2}))/u_\Delta\label{eq:vDeltaF2}
\end{align}
\end{subequations}
The complete prescription for solving the system of equations then involves: solving for the outbreak size of the disease-1 using Eqs \ref {eq:disease1G_1} and \ref{eq:disease1G_0} before solving the simultaneous system of Eqs \ref{eq:vbotA} to \ref{eq:vDeltaF2} to find the $v$-values and finally using Eq \ref{eq:disease2G0} to find $S_2$. The fraction of the network that then contracts both diseases is then given by $S_1(1-S_2)$.
\begin{figure}[ht!]
\begin{center}
\includegraphics[width=0.45\textwidth]{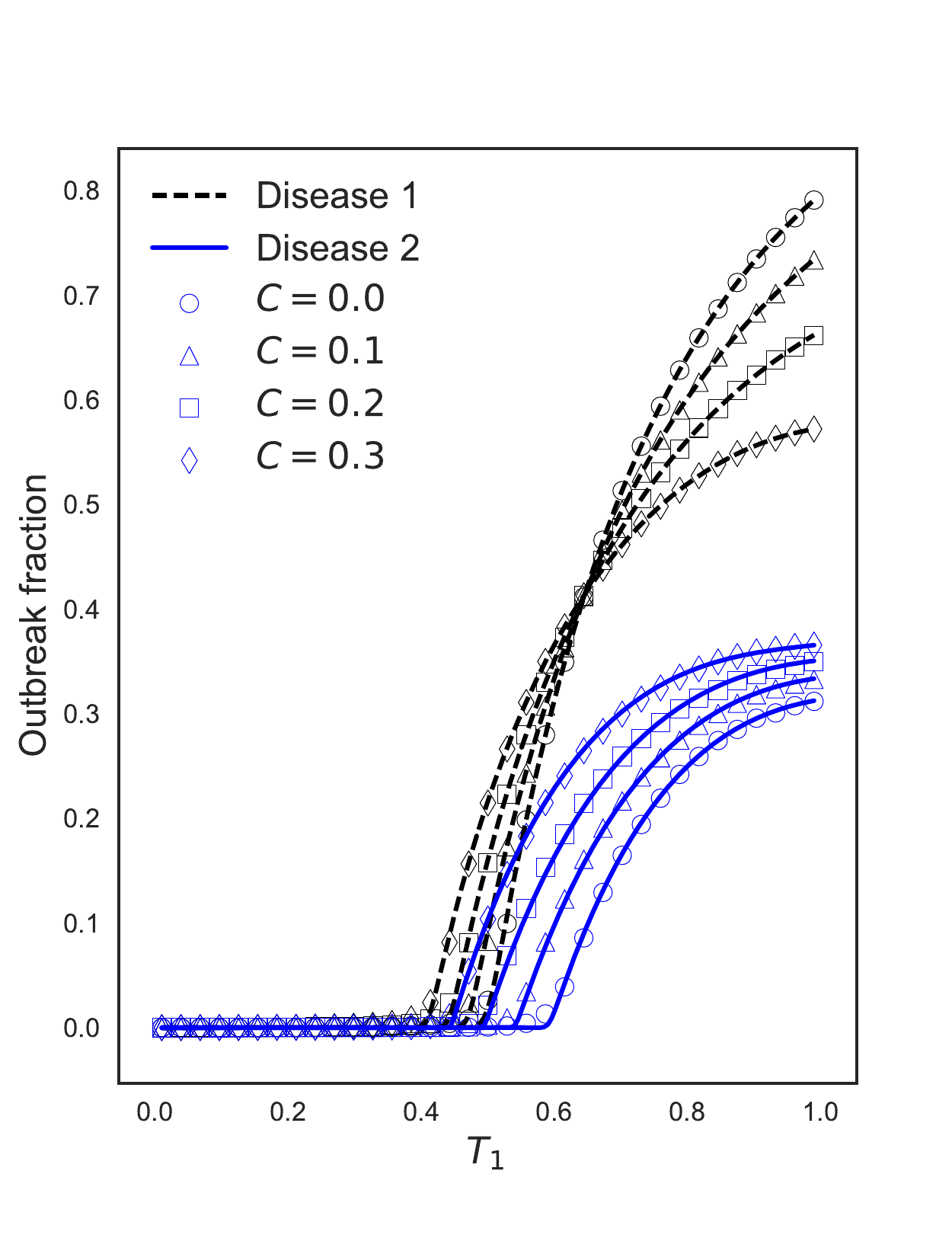}
\caption[main]{ 
The outbreak fractions for the 2-strain coinfection model on ER random graphs with $T_2=0.6$, $N = 25000$, for varying clustering coefficients and fixed mean degree $\langle s\rangle+2\langle t \rangle=2$. It is clear that clustering reduces the epidemic thresholds of both diseases and reduces the outbreak size of the first strain; however, it \textit{increases} the coinfected fraction of the network. Points are the average of 500 repeats of Monte Carlo simulation while solid lines are the theoretical predictions from Eqs \ref{eq:disease1G_0} and \ref{eq:disease2G0}.
} \label{fig:main}
\end{center}
\end{figure}

\section{Discussion}
\label{sec:discussion}

\subsection{The double Poisson distribution}
\label{subsec:dpd}

\begin{figure}[ht!]
\begin{center}
\includegraphics[width=0.45\textwidth]{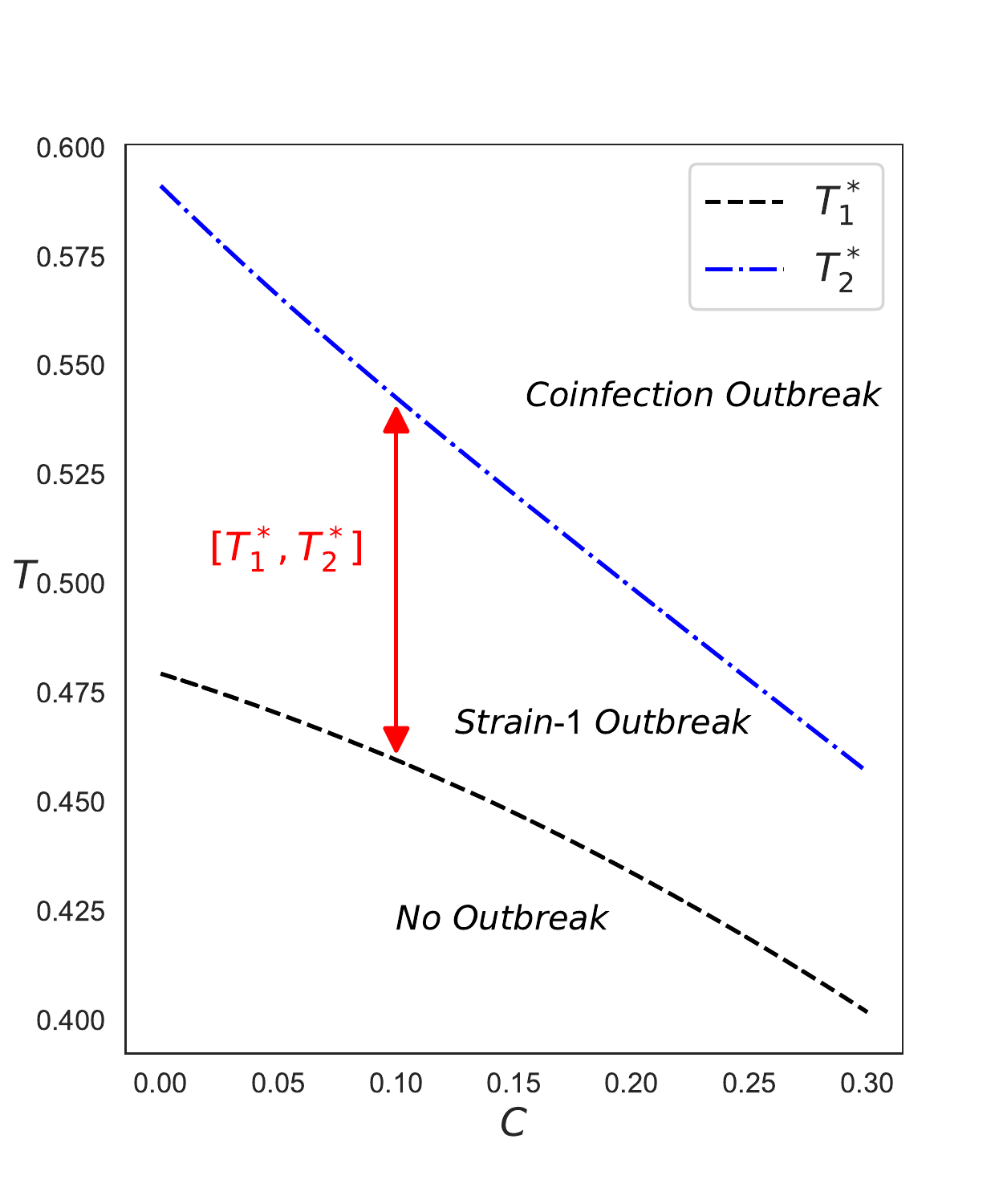}
\caption{ An analytical investigation of the critical transmissibilities of both strains for increasing clustering coefficients for the ER networks defined in Fig \ref{fig:main}. The curves are generated from Eqs \ref{eq:disease1G_0} and \ref{eq:disease2G0} at the value of $T$ in which the outbreak fraction becomes large than $\epsilon$. The critical points reduce as $C$ increases, indicating that contact clustering helps to spread an emergent epidemic. The interval $[T_1^*,T_2^*]$ is the transmissibility window in which strain-1 exists as an epidemic on the network without strain-2. We observe that increased clustering reduces the interval and thus increases the extent of coinfection in this model, at the expense of the mono-infection equilibrium. We note, however, that clustering can never reduce the coinfection critical point below the single strain threshold due to the strict conditions of perfect coinfection in the premise of the model.   
} \label{fig:R0}
\end{center}
\end{figure}

An example of the 2-strain model is shown in Fig \ref{fig:main} for ER networks with fixed mean degree $\langle k \rangle=\langle s\rangle +2\langle t\rangle=2$. In the limit of large network sizes, the joint degree distribution is given by the double Poisson distribution
\begin{equation}
    p(s,t) = e^{-\langle s\rangle}\frac{\langle s\rangle ^s}{s!}e^{-\langle t\rangle}\frac{\langle t\rangle ^t}{t!}
\end{equation}
In this instance, the generating functions $G_0(x,y)=G_{1,\bot}(x,y) = G_{1,\Delta}(x,y):= G(x,y)$ and we can write 
\begin{equation}
    G(x,y) = e^{\langle s\rangle(x-1)}e^{\langle t\rangle(y-1)}
\end{equation}
The clustering coefficient for these graphs is then found using Eq \ref{eq:clusteringCoeff} to be 
\begin{equation}
    C = \frac{2\langle t\rangle}{2\langle t\rangle +\langle k\rangle^2}
\end{equation}
We can use this expression to determine the tree degree and the average number of triangles per node for a fixed average degree, $\langle k\rangle$, along with $\langle k \rangle = \langle s\rangle+2\langle t\rangle$. We use this expression in Fig \ref{fig:R0} to numerically investigate the epidemic thresholds of the two strains as a function of increasing clustering coefficient. An outbreak is considered to be an epidemic if the fraction of the network infected, $S(T_1)$, is larger than $\epsilon = 1\times 10^{-3}$, and hence, we can use Eqs \ref{eq:disease1G_0} and \ref{eq:disease2G0} to approximate the epidemic threshold of each strain to sufficient accuracy. It is clear that increased clustering coefficients lead to a broadening of the region of the model's parameter space which can sustain coinfection at the increasing expense of the single strain equilibrium. 

\subsection{Clustered human contact networks}
\label{subsec:ocm}

The double Poisson model can display the properties of clustering as well as afford an exact solution. However, it is known that the distributions of contacts in many social networks are not well represented by ER graphs; instead, they often follow a power law. 
Further, whilst the average degree of ER networks is fixed, the degree correlations are known to change with increasing $C$, which has been the subject of much research \cite{miller_2009,PhysRevE.101.062310,PhysRevE.81.066114}. In this section, we investigate another clustering model that is more representative of real-world social networks as well as holding the degree correlations steady as we vary the clustering coefficient.

\begin{figure}[ht!]
\begin{center}
\includegraphics[width=0.50\textwidth]{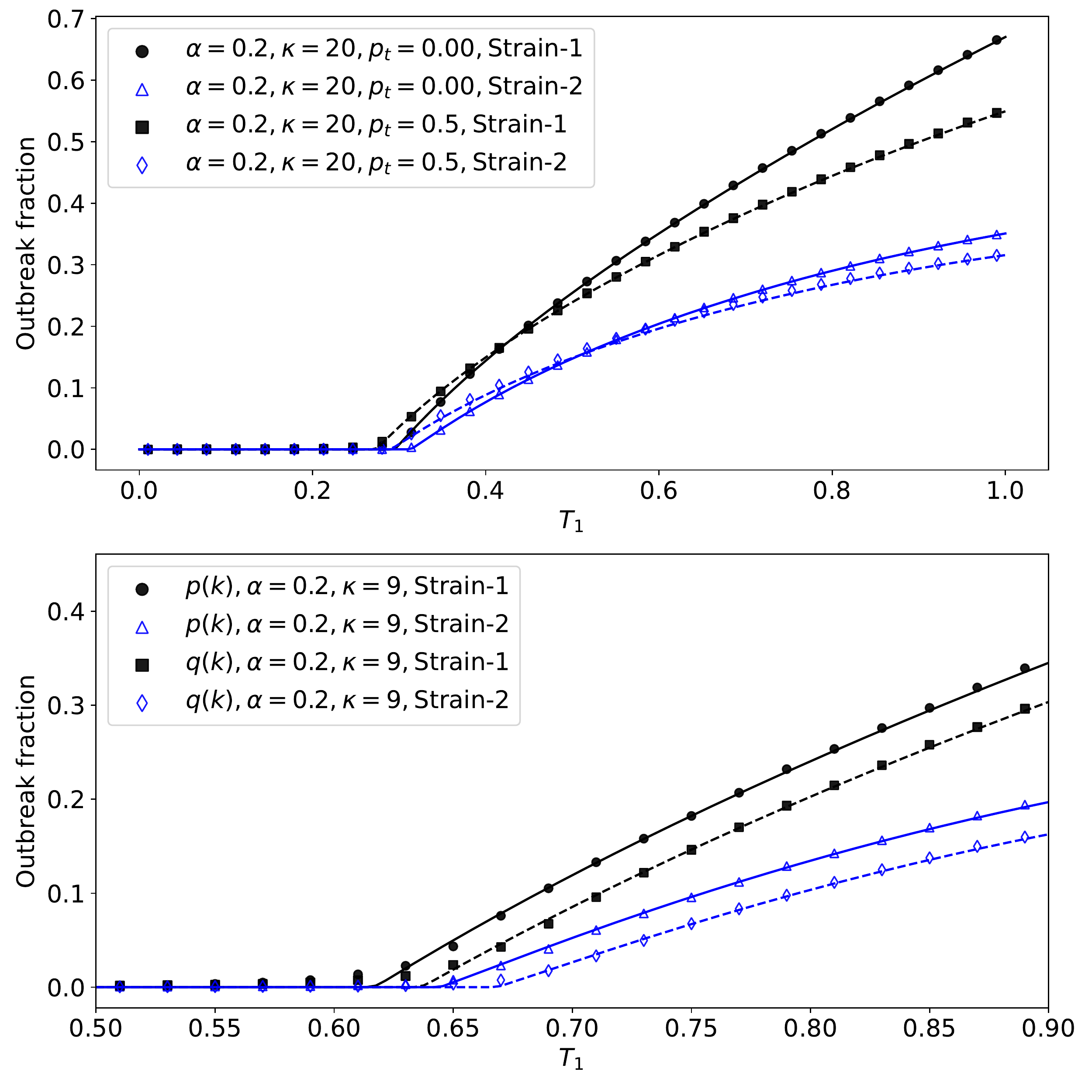}
\caption{ (Top) The outbreak fractions of both strains for increasing triangle probabilities within the CHCN degree distribution defined in Eq \ref{eq:CHCN}. Clustering reduces the epidemic threshold, in agreement with ER experiments; however, coinfection \textit{decreases} with increasing $\theta$. (Bottom) The expected epidemic sizes of both strains for a $p^{\text{CHCN}}(k)$ network with $\theta=0.0$ and a degree-$\delta$ CHCN distribution network defined by Eq \ref{eq:CHCNQ}. Clustering \textit{increases} the epidemic threshold in this model. Markers are the average of 100 repetitions of bond percolation on CHCN networks with $N=1e^5$ nodes and $T_2=0.6$.
} \label{fig:CHCN}
\end{center}
\end{figure}

We propose a realistic model of human contact networks with tunable clustering based on Newman \cite{PhysRevE.66.016128} and Hackett \textit{et alia} \cite{PhysRevE.83.056107,PhysRevE.81.066114}. Contact networks often follow a scale-free (SF) distribution and of particular importance is the SF degree distribution, $p^{\text{SFC}}(k)$, whose maximum degree is curtailed by an exponential degree cut-off (SFC). Such a distribution is given by 
\begin{equation}
    p^{\text{SFC}}(k) = ck^{-\alpha}e^{-k/\kappa}
\end{equation}
where $2\leq \alpha\leq 3$ is the power law exponent, $\kappa\in \mathbb Z$ is the degree cut-off and $c$ is a normalisation constant. In order to extend this degree distribution to the tree-triangle model, it is necessary to decompose the degree of a node, $k$, into tree-like and triangle contributions. We achieve this by introducing $\theta$ as the probability that a given node is a member of precisely $t$ triangles. Thus, the clustered human contact network (CHCN) has tree-like and triangle degrees distributed according to 
\begin{equation}
    p^{\text{CHCN}}(k) = p^{\text{SFC}}(k) \sum_{t=0}^{\lfloor k/2\rfloor}\binom{\lfloor k/2\rfloor}{t}\theta^{t}(1-\theta)^{\lfloor k/2\rfloor-t}\label{eq:CHCN}
\end{equation}
where $\lfloor \cdot \rfloor$ is the floor function. The normalisation constant can be found from the condition 
\begin{equation}
    \sum_{k=0}^\infty p^{\text{CHCN}}(k)=1
\end{equation}
Configuration model networks generated using this distribution have an identical distribution of overall degrees; however, their tree-like and triangle decompositions are modulated through $\theta$. For high $\theta$ values, the heavy tail of the distribution is able to introduce a significant amount of clustering into the network. We observe the percolation properties of the model for increasing $\theta$ in the top plot of Fig \ref{fig:CHCN}. We find that clustering reduces the epidemic threshold of both strains and, in contrast to the ER model, \textit{reduces} the amount of coinfection in the network. 
Whilst CHCN networks have identical overall degrees, there is no control of the degree assortativity across the different experiments, however. Further, when the degree cut-off is large, we expect minimal assortativity, especially among the numerous low-degree sites. To examine the effect of assortativity on these results, we propose a use the degree-$\delta$ model \cite{miller_2009} to the distribution to allow control of the degree correlations. The degree distribution is given by
\begin{equation}
    p^{\text{CHCN}, \delta}(k) = \begin{cases}
    p^{\text{CHCN}}(k)\delta_{k,s}\delta_{t,0}  \qquad k\neq 3\\
    p^{\text{CHCN}}(k)\delta_{s,1}\delta_{t,1}\qquad k=3
    \end{cases}\label{eq:CHCNQ}
\end{equation}
In other words, nodes in the network are not clustered unless their overall degree is $k=3$, in which case, they are involved in exactly one triangle and one independent edge. This distribution forces the clustering to remain among the low-degree nodes towards the periphery of the network and thus, we expect clustering to be positively assorted. We examine this degree distribution in the bottom plot of Fig \ref{fig:CHCN} for a lower degree cut-off and find, in agreement with \cite{miller_2009,PhysRevE.81.066114,PhysRevE.101.062310}, and in contrast to the results from Eq \ref{eq:CHCN}, that clustering \textit{increases} the epidemic threshold of both strains.

\section{Conclusion}
\label{sec:conclusion}

In this paper we have studied the effects of clustering on the spread of two diseases that interact via perfect coinfection. Under the bond percolation isomorphism, this scenario considers a bond percolation process over the GCC of a previously percolated graph with clustering. The presence of triangles in the substrate network creates non-trivial network motifs that subsequent percolation processes experience as they spread over the network. The fracturing of the triangles by the first percolation process leads to a loss of symmetry for these subgraphs, causing their $f_\Delta$ expressions to become complicated for the subsequent percolating process. 
We derived the conditions for the expected outbreak size of each of the diseases and found that clustering decreases the epidemic threshold for both diseases, for two disassortative configuration model random graphs, ER and CHCN (in the limit of large degrees).

For ER networks, clustering decreases the outbreak size at high transmissibilities for the first disease; however, it \textit{increases} the infected fraction of the second strain at a given transmissibility. We also find that clustering reduces the width of the interval between the epidemic thresholds $[T_1^*,T_2^*]$, indicating, intuitively, that coinfection occurs more readily when contacts are clustered together in ER networks.

Conversely, we showed that clustering reduces the extent of coinfection for high-degree CHCN networks. However, when low-degree and positively assorted CHCN networks were considered, clustering was found to increase the epidemic threshold of both strains and reduce the extent of coinfection present at equilibrium. This final example is thought to well represent the effects of social distancing, a population management strategy designed to isolate individuals into contact bubbles. Our model replaces $k=3$, well mixed nodes, by a peripheral, outer-core structure comprised of predominantly low-degree nodes. Within this context, social distancing is shown to be an excellent mechanism to reduce disease spreading among a population. However, further investigation is required to understand the precise role of $\kappa$ and assortativity for transitioning between the percolation properties of the distribution in Eq \ref{eq:CHCN} and those of Eq \ref{eq:CHCNQ}, which we leave for future work. 

We conclude, therefore, that the effects of clustering on the extent of coinfection among a populace are highly dependent on the details and arrangement of the distribution of triangles in the network. We have shown coinfection can be increased or decreased with clustering as well as moving the epidemic threshold of the second strain to higher and lower critical points. 

The model presented here can be generalised in numerous ways, including the consideration of higher-order cycles in the substrate contact network (and thus describing many social networks and community structures) or, through the incorporation of subsequent diseases over the increasingly rich immunological landscape. We imagine that these results will be useful to the mitigation of coinfections and to disease-prevention strategies among human contact networks.

\subsection*{References}

\bibliography{bib}

\end{document}